\documentclass[12pt]{article}
\providecommand{\keywords}[1]{\textbf{Keywords:} #1}
\usepackage{setspace}
\usepackage[a4paper,margin=1.2in]{geometry}
\usepackage[utf8]{inputenc} 
\usepackage[T1]{fontenc}    

\usepackage{authblk}        
\usepackage{graphicx} 
\usepackage{multirow}%
\usepackage{amsmath,amssymb,amsfonts}%
\usepackage{amsthm}%
\usepackage{mathrsfs}%
\usepackage[title]{appendix}%
\usepackage{xcolor}%
\usepackage{textcomp}%
\usepackage{manyfoot}%
\usepackage{booktabs}%
\usepackage{algorithm}%
\usepackage{algorithmicx}%
\usepackage{algpseudocode}%
\usepackage{listings}%
\usepackage{natbib}
\usepackage{caption}
\usepackage{float}
\usepackage{array}

\usepackage{xr-hyper} 
\usepackage{url} 
\usepackage[hidelinks]{hyperref} 
\usepackage{cleveref}       
\usepackage{adjustbox}
\usepackage{nameref}
\usepackage{mathtools}
\usepackage{doi}
\usepackage{makecell}
\usepackage{comment}

\title{Sustainable Development Goals in Psychology: A Century of Progress in Publications}
\author{Xinyi Zhao$^1$, Ralph Hertwig$^1$, Dirk U. Wulff$^{1,2}$}
\affil{$^1$Center for Adaptive Rationality, Max Planck Institute for Human Development\\$^2$University of Basel}

\begin{document}
\maketitle

\abstract{\noindent The Sustainable Development Goals (SDGs) offer a lens for tracking societal change, yet contributions from the social and behavioral sciences have rarely been integrated into policy agendas. To take stock and create a baseline and benchmark for the future, we assemble 233,061 psychology publications (1894–2022) and tag them to the 17 SDGs using a query-based classifier. Health, education, work, inequality, and gender dominate the study of SDGs in psychology, shifting from an early focus on work to education and inequality, and since the 1960s, health. United States-based research leads across most goals. Other countries set distinct priorities (e.g., China: education and work; Australia: health). Women comprise about one-third of authors, concentrated in social and health goals, but have been underrepresented in STEM-oriented goals. The 2015 launch of the SDGs marked a turning point: SDG-tagged publications have been receiving more citations than comparable non-SDG work, reversing a pre-2015 deficit. Tracking the SDGs through psychology clarifies long-run engagement with social priorities, identifies evidence gaps, and guides priorities to accelerate the field’s contribution to the SDG agenda.}

\noindent\\ \keywords{sustainable development goals, SDGs, bibliometrics, natural language processing, psychology}

\section*{Introduction}

During the past century, rapid technological, economic, and social transformations have driven notable global progress, but these transformations have also exacerbated challenges such as climate change, environmental degradation, public health crises, and social inequality \citep{Malekpour_SDG_Scientist_2023}. To confront these interconnected issues, the United Nations adopted the Sustainable Development Goals (SDGs) in 2015 as a universal framework to promote human well-being, equity, and environmental sustainability. Central to this agenda are the 17 Sustainable Development Goals (SDGs), which call for collective action from all scientific disciplines and sectors of society to address pressing challenges, including poverty, inequality, health, education, economic growth, and climate change \citep{UN_SDGs}. At the same time, achieving these goals requires understanding and changing human behaviors, motivations, and collective dynamics that facilitate or impede progress. Psychology offers essential insights in this regard. In this study, we map the historical trajectories of SDG-related research in psychology and identify patterns of engagement with goals and populations.

Psychology contributes an essential perspective to the SDG agenda, as human behavior and behavior change are integral to nearly every SDG \citep{hertwig2025moving, nielsen2024realizing}. Mental health, a key dimension of Good Health and Well-being (SDG 3), illustrates the direct relevance of psychology: psychological science has produced evidence-based interventions, such as cognitive–behavioral therapy, resilience training, and community-based programs that improve mental health care worldwide \citep{Patel_mental_2018}. In the domain of Quality Education (SDG 4), educational and social psychology investigate how motivation, identity, and social belonging shape learning outcomes and educational inequalities \citep{Easterbrook_education_2020}. In addition, research on how people make decisions, respond to incentives, and navigate uncertainty provides foundational information for many global challenges and behavior change. Behavioral public policy, judgment and decision-making, and research on risk perception, behavior, and communication, often bridging psychology with economics, political science, and public policy, show how human behavior shapes social and environmental outcomes across diverse SDGs, such as Decent Work and Economic Growth (SDG 8), Reduced Inequalities (SDG 10), and Peace, Justice and Strong Institutions (SDG 16). However, despite this broad relevance, psychology’s participation in the SDGs has not been systematically analyzed. Understanding how psychological research aligns with the SDGs is therefore crucial for evaluating how human behavior, cognition, and social processes contribute to social change and for establishing a baseline from which future developments can be assessed.

Tracking SDG-related scientific research reveals not only how global challenges evolve over time, but also who participates in addressing them \citep{Malekpour_SDG_Scientist_2023, BarberMarin_biblioSDG_Euro_2024}. Understanding participation patterns is crucial for identifying inequalities in global knowledge production and ensuring that a diversity of voices contributes to sustainable solutions. National priorities for SDG research often reflect differences in developmental trajectories, policy agendas, and scientific capacities \citep{Kraemer_country_2022}. Examining such cross-national variations helps contextualize global research trends and identify areas where progress remains uneven. A gender lens adds another crucial dimension. Women and men may contribute differently to SDG-related research, both in their representation and in the thematic areas they emphasize \citep{Huang_gender_science_2020, Beloskar_gender_SDG5_2023}. Examining these gendered dynamics not only reveals disparities in scientific participation but also helps identify how diversity in the research community may shape the framing of global challenges and approaches to their solutions. 

Bibliometric analysis provides a systematic approach to understanding how SDG-related research evolves over time and how different research communities engage with it. By examining patterns of publication and authorship, such analyses reveal who contributes to SDG research, across countries, institutions, and genders, and how these contributions have changed over time \citep{Bornmann_SDGMapping_2023}. Bibliometric indicators such as citation counts further capture the level of attention that SDG-related studies receive within the scientific community, offering insight into their visibility and perceived impact \citep{Ottaviani_SDGclassification_2024}. Together, these perspectives enable a more comprehensive evaluation of global knowledge production on sustainable development and point to factors that can enhance the recognition and influence of SDG-oriented research.

A key methodological challenge in assessing psychology’s contribution to the SDGs using bibliometric data is determining which publications are SDG-related. Traditional bibliometric analyses typically rely on narrow keyword searches, for instance, to identify publications that directly refer to the SDGs \citep{Aftab_healthSDG_2020, Yamaguchi_biblioSDG_2022, BarberMarin_biblioSDG_Euro_2024}, potentially overlooking research that contributes to the same issues without using narrow SDG terminology. As the SDGs were only formally adopted in 2015, this approach also biases the sample toward more recent publications, underrepresenting decades of earlier research that laid the scientific foundations for sustainable development. However, progress on issues such as poverty, climate change, health, education, and gender equality precedes the SDGs by decades. To address these limitations, broader query-based labeling systems have been developed that enable a more comprehensive, historically grounded assessment of how science supports the SDG agenda \citep{Wulff_SDG_2024}.

To provide a global and historical perspective on the evolution of SDG-related research, this study adopts a pioneering approach employing text2sdg, an ensemble system based on multiple query-based SDG labeling systems \citep{text2sdg, Wulff_SDG_2024}, to identify SDG themes within 233,061 psychology publications spanning from 1894 to the present, collected by the American Psychological Association (APA), the leading publisher in the field of psychology. Our objective is to uncover overarching trends, investigating how psychological research has engaged with SDG-related areas and how it has evolved over time. We further examine diversity and inclusivity within SDG-related psychological research, with particular attention to national contributions and authorship patterns. We also investigate the scientific impact of these publications, thus providing insight into the visibility and influence of psychology’s involvement with the SDGs. This study provides a systematic account of how psychology has engaged with the SDGs to date, establishing an empirical foundation to understand how the priorities of the discipline have evolved over time and providing a baseline against which future developments can be assessed.

\begin{figure}[htbp!]
\centering
\includegraphics[width=1\textwidth]{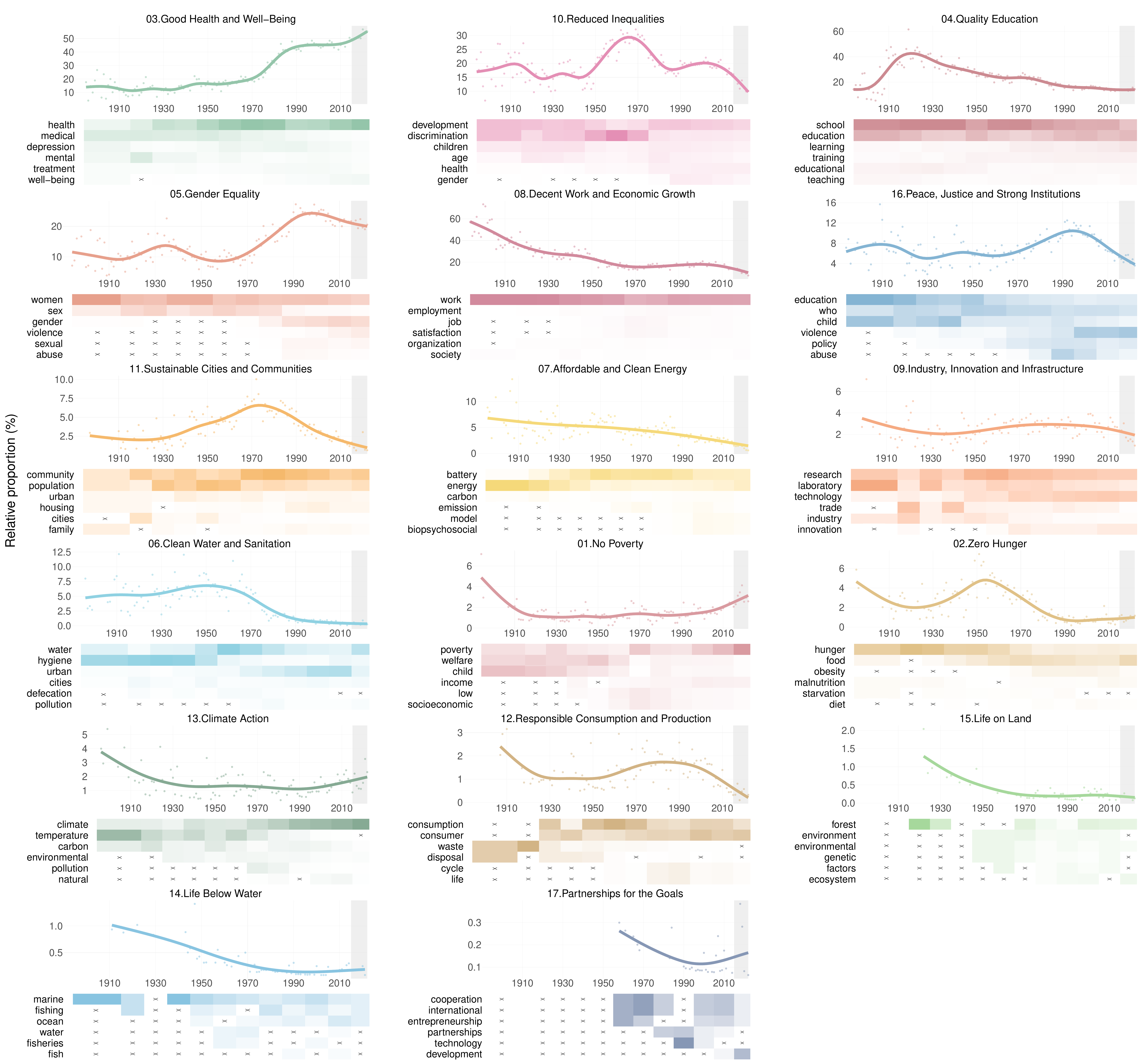}
\caption{\textbf{Trends in SDG-related psychological research and associated topics over time.} Panels are ordered by their overall proportion among all psychology publications. The x-axis indicates publication year, and the y-axis indicates the proportion of psychology publications associated with each SDG among all SDG-related papers. In each SDG panel, the three most frequent keywords are shown for the historical periods pre-1920, 1920–1959, 1960–1999, 2000–2014, and post-2015, with percentages in parentheses indicating their relative frequency.}\label{fig:sdg_trend}
\end{figure}

\section*{Results}

Our results are structured in three parts. We first examine historical and long-term trends in SDG-related research within psychology, tracing how attention to different goals has evolved over time and across thematic domains. We then analyze national and gendered patterns of engagement, identifying how research contributions vary between countries and between male and female researchers. Finally, we assess the scientific impact of SDG-related publications through citation analyses, evaluating if and to what extent association with the SDGs is linked to greater scholarly visibility and influence.

\subsection*{Trends in SDG-related Psychological Research}

Overall, the prominence of SDG-related psychology, although formally articulated only in 2015, has been evident for more than a century. The proportion of psychology publications addressing any of the SDGs has fluctuated substantially over time, beginning to rise at the end of the nineteenth century and reaching nearly 40\% before 1920. After a brief decline, attention to SDG-related topics increased steadily again, peaking at around 35–40\% in the 1990s. See Supplementary Information, Fig. S1. Although SDG-linked research has continued to increase in absolute terms, the relative share of SDG-related publications has shown a downward trend, including in the years following 2015. Topic-wise, psychologists’ engagement with the SDGs has evolved unevenly over time. A small set of goals, most notably Good Health and Well-being (SDG 3), Reduced Inequalities (SDG 10), Quality Education (SDG 4), Gender Equality (SDG 5) and Decent Work and Economic Growth (SDG 8), represent the majority of psychological research related to the SDGs in general (89.40\%), although their relative prominence has changed over historical periods. In earlier decades, psychological research was heavily oriented toward education and employment, whereas since the 1970s, health-related topics have grown substantially in importance. By contrast, goals related to environmental sustainability, such as Climate Action (SDG 13), Life on Land (SDG 15), and Life Below Water (SDG 14), have remained consistently underrepresented, accounting for only 1.6\% of psychological publications related to the SDGs.
 
Fig. \ref{fig:sdg_trend} displays the trends in SDG-related psychological research for each goal. More specifically, Decent Work and Economic Growth (SDG 8) was the earliest dominant theme, especially before 1920, when roughly 40\% of SDG-linked psychology publications were work-related. This emphasis on employment and organization coincided with the disruptions of industrialization and the Great Depression, extending beyond economics into psychological studies of labor and productivity. From the 1920s to the 1960s, attention moved toward Quality Education (SDG 4), though it fell from 38\% to 25\%, and Reduced Inequalities (SDG 10), which rose from 20\% to 27\%. It reflects a growing concern for universal education and a subsequent shift toward social inequity. Since the 1960s, Good Health and Well-being (SDG 3) has risen steadily, from approximately 20\% to more than half after the adoption of the SDGs in 2015, underscoring psychology’s growing role in physical and mental health research. In contrast, environmentally oriented goals such as Climate Action (SDG 13), Life Below Water (SDG 14), and Life on Land (SDG 15) remain marginal, each accounting for less than 1\% of publications.

Even within the most prominent SDGs, the substantive focus of research has evolved. For example, health research shifted from terms such as “medical” and “disease” to “depression,” “well-being,” and “suicide,” indicating the societal recognition of the importance of psychological health and well-being as central to sustainable human development. Similarly, gender-related research moved from “women” and “sex” to “gender” and “violence,” reflecting the growing influence of feminist movements, the recognition of gender as a social construct, and an increased awareness of structural and interpersonal inequalities. In education, the centrality of “school” has been complemented by “learning,” aligned with pedagogical reforms, the rise of cognitive psychology and later the digital transformation of education. More broadly, these thematic shifts underscore how psychology has adapted to changing social priorities: from addressing basic survival and institutional access to engaging with complex, multifaceted challenges of equity, mental health, and human development in an interconnected world.

In general, each SDG that resonates with major social issues has deep historical roots in psychology, reflecting the field's long-term involvement in health, education, work, inequality, and gender. Recognizing these continuities reveals that psychology as a research endeavor has long mirrored social priorities and public attention. By following these trajectories, we gain not only a clearer picture of the evolving focus of the discipline but also a broader understanding of how psychological research is shaped by historical transformations and social change.

\begin{figure}[htbp!]
\centering
\includegraphics[width=1.05\textwidth]{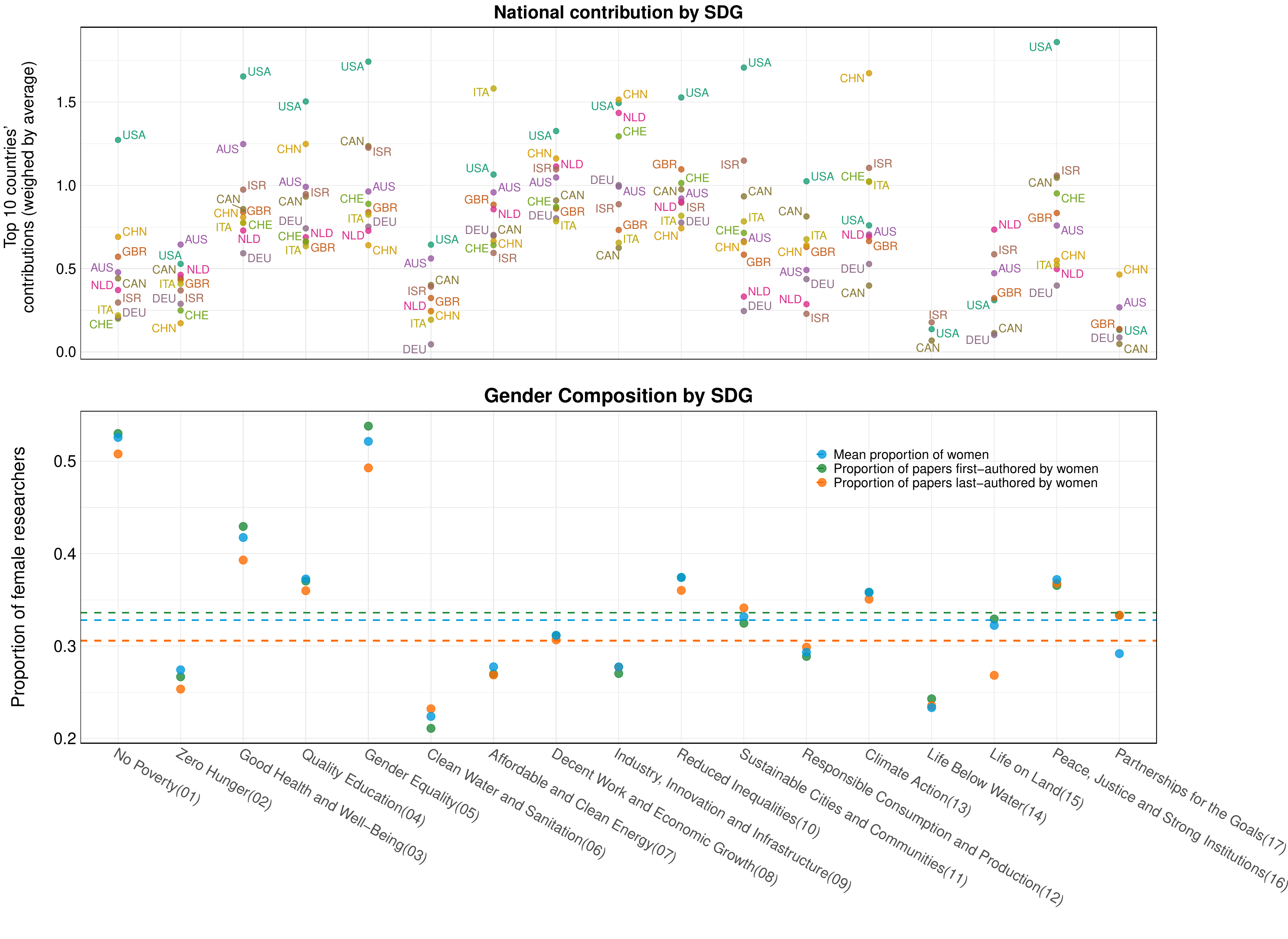}
\caption{\textbf{National and gendered contributions to SDG-related psychological research.} Panel (a) compares the country-level contributions to SDG-related psychological research among the top ten countries with the largest overall output of SDG-related psychology publications (USA, CAN, GBR, AUS, DEU, NLD, ISR, CHE, ITA, CHN) across all SDGs. The x-axis represents the SDGs, and the y-axis shows each country’s percentage share of publications within that SDG relative to the overall share of publications associated with that SDG. Panel (b) presents the mean proportion of women authors per publication, and the shares of publications first- and last-authored by women, across each SDG. The dotted lines indicate the average level of each corresponding proportion for all SDGs.}\label{fig:sdg_ctr_gender}
\end{figure}

\subsection*{National and Gendered Contributions to SDG-related Research }

Although psychology contributes to many SDG-related domains, little is known about which countries and researchers are driving this engagement. Examining national and gender patterns offers a clearer picture of how participation in SDG-related psychological research is distributed globally and whose contributions shape the field.

In assessing national contributions to SDG-related psychological research, we restricted the comparison to publications after 1988 (87.7\% of all records), as country information is incomplete for previous years. Because many publications involve authors from multiple countries, national shares represent the proportion of publications with at least one author affiliated with that country. The United States (U.S.) alone produces an exceptionally large share of SDG-related publications (72.6\%), far exceeding Canada (6.0\%), the United Kingdom (3.8\%), Australia (2.4\%) and Germany (2.4\%), followed by the Netherlands (1.5\%), Israel (1.2\%), Switzerland (0.7\%), Italy (0.6\%), and China (0.6\%).

Fig.~\ref{fig:sdg_ctr_gender} (a) shows, for the top ten contributing countries, the share of psychology publications related to each SDG relative to the global average. Values above 1 indicate that a country contributes more to that SDG than expected based on its overall publication share. The results reveal that the U.S. not only dominates in total output, but also contributes disproportionately to multiple SDGs. 

Other top-contributing countries show more selective engagement with specific SDGs. Canada contributes relatively more to Gender Equality (SDG 5) and Peace, Justice and Strong Institutions (SDG 16), while Australia’s output is more concentrated in health-related research, particularly Good Health and Well-being (SDG 3). The United Kingdom and Germany, despite being among the most productive countries overall, place comparatively less emphasis on psychological research focused on the SDGs. The United Kingdom shows engagement close to the global average for Reduced Inequalities (SDG 10), whereas Germany demonstrates a similar, around-average focus on Industry, Innovation and Infrastructure (SDG 9). 

China, amid rapid development, shows stronger representation in Quality Education (SDG 4), Decent Work and Economic Growth (SDG 8), and Climate Action (SDG 13), but contributes less to Gender Equality (SDG 5) and Reduced Inequalities (SDG 10). Italy emphasizes Affordable and Clean Energy (SDG 7) but, unlike China, contributes less to economy-oriented goals such as Decent Work and Economic Growth (SDG 8) and Industry, Innovation and Infrastructure (SDG 9). 

In addition to these national differences, the gendered authorship patterns shown in Fig.~\ref{fig:sdg_ctr_gender} (b) indicate varied participation in the SDGs. In general, women make up roughly one-third of all authors in SDG-related psychology publications, with similar proportions among the first authors and slightly lower representation among the last authors. However, their involvement differs substantially between thematic areas. Women are more strongly represented in socially and health-oriented SDGs, particularly Gender Equality (SDG 5) and No Poverty (SDG 1), where they account for more than half of the first authors and, in the case of SDG 5, also the last authors. In contrast, their participation is lower in STEM-related SDGs, such as Affordable and Clean Energy (SDG 7) and Life Below Water (SDG 14), reflecting psychology's broader disciplinary focus in its engagement with the SDG agenda.

Taken together, SDG-related research in psychology shows substantial variation between countries and between genders. Differences in national research priorities and gender representation patterns shape the specific areas of the SDG agenda.

\begin{figure}[htbp!]
\centering
\includegraphics[width=1.05\textwidth]{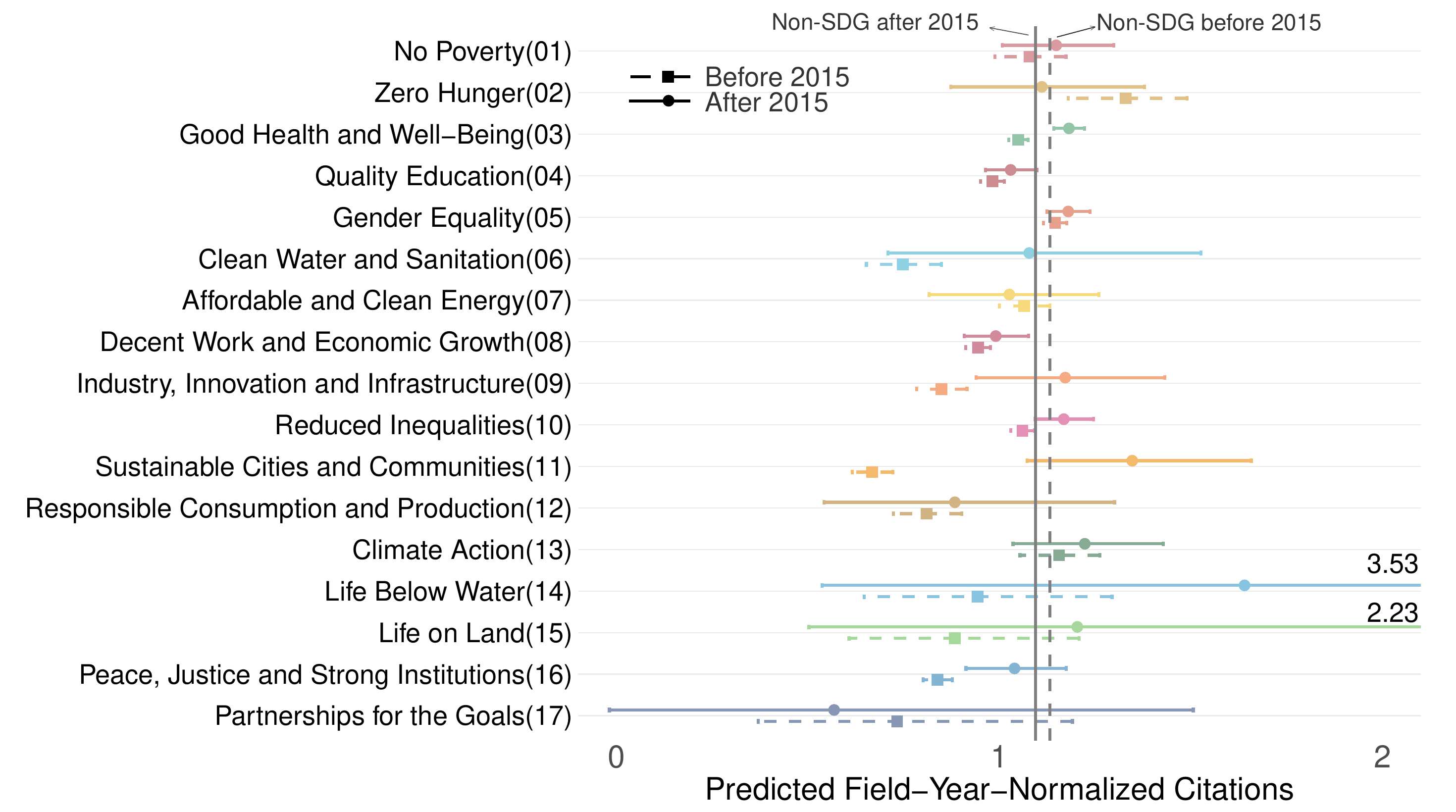}

\caption{\textbf{predicted citation scores of SDG papers before versus after 2015} The dashed vertical line marks the baseline citation level for non-SDG publications before 2015, while the solid vertical line marks the baseline after 2015. Empty circles with dashed 95\% confidence intervals show the estimated citation counts for SDG-related publications in that goal prior to 2015, whereas solid circles with solid confidence intervals show the estimates for SDG-related publications after 2015}\label{fig:sdg_citation}
\end{figure}

\subsection*{Citation patterns of SDG-related publications in Psychology}

To investigate whether publications associated with the SDGs received greater scholarly attention, we use a difference-in-differences (DiD) specification. These citation scores were normalized within psychology subfields and publication years, allowing us to compare the citation impact across the entire body of publications. Because the SDGs were formally introduced in 2015, we expected that publications explicitly linked to the SDGs after this point might receive more attention than earlier work that addressed related issues without explicit SDG framing. To test this, we include the interaction between the SDG association and the publication period (before versus after 2015). In addition, our models account for several factors known to influence citation impact, including the number of authors, the gender composition of the author team (first author female, last author female, and overall proportion of female authors), the affiliated countries (dummy variables for the top ten countries), and the impact factor of the publishing journal. After excluding records with missing information on these variables and publications without citation data (due to missing DOIs), the analytic sample consisted of 180,471 publications (77.4\% of the original dataset). The normalized citation scores remained highly right-skewed (mean = 1.10, s.d. = 2.36), so we applied a log transformation to stabilize the distribution prior to modeling.

Based on the DiD model, Fig.~\ref{fig:sdg_citation} presents the predicted normalized citation counts by SDG and period, holding other covariates constant. The dashed and solid lines indicate the benchmarks for non-SDG publications before and after 2015, respectively. Compared to this baseline, non-SDG publications experienced a decrease in citations after 2015, whereas several SDG-related topics gained citation visibility over the same period. Among the control variables, the journal impact factor and team size were positively associated with citation impact, while gender composition showed little effect. Full model estimates are reported in Supplementary Information (SI), Table~S1.

Good Health and Well-being (SDG 3), the most studied goal in psychology, exhibits a significant post-2015 increase, with citation counts exceeding those of non-SDG publications. Notably, SDG 3-related psychology publications had received fewer citations than non-SDG work before 2015, suggesting a reversal in citation patterns following the introduction of the SDG framework. Other goals also display significant gains, including Sustainable Cities and Communities (SDG 11), Industry, Innovation and Infrastructure (SDG 9), and Peace, Justice and Strong Institutions (SDG 16). In particular, SDG 11 publications increased from about 0.67 normalized citations before 2015 to 1.35 after 2015, surpassing the level of comparable non-SDG work. Although SDG 16 also shows a significant improvement, its citation levels remain closer to those of non-SDG publications. These results indicate that the explicit SDG framework has improved the scholarly visibility of psychological research, although unevenly between goals.

\section*{Discussion}

This study provides the first comprehensive, long-term analysis of psychology’s contributions to the SDGs, drawing on more than 233,000 publications from 1894 to 2022. By systematically tagging psychological research to the 17 SDGs, we uncover shifting emphases from the early focus on work and education to the growing prominence of health, inequality and gender, shaped by broader societal transformations and national contexts. The global distribution of SDG-related psychological research is dominated by the U.S. and a small group of Global North countries, such as Canada and the UK, while emerging contributions from regions, including China, reflect the field's more recent diversification. Women account for roughly one-third of authors, with greater participation in health- and inequality-related SDGs but lower participation in STEM-oriented areas. The launch of the SDGs in 2015 marked a turning point: psychology publications linked to the goals thereafter received more citations than comparable non-SDG work, underscoring their rising visibility and recognition. These findings highlight psychology’s important but uneven role in advancing sustainable development. It underscores the need to integrate behavioral and social insights more fully into global policy agendas.

Viewed through a social- and behavioral-science lens, SDG-related research functions as a barometer of public policy priorities: attention rises as problems become salient in everyday life and policy \citep{Sarewitz_science_2016}. Psychology’s focus has changed with societal change, from work and organization during late industrialization to a mid-century pivot toward education and inequality amid universal schooling and civil rights reforms, and since the 1960s, a sustained rise in mental health and populations with medicalization and destigmatization \citep{Pickren_psychology_2010}. However, the way we frame mechanisms and interventions partly reflects who participates \citep{Harding_diversity_2015}. SDG-related psychological research is disproportionately driven by well-funded systems, especially the U.S., which are better positioned to translate scientific evidence into policy \citep{Ciarli_globalscience_2021, Tijssen_societalImpact_2016, wagner2017opencountries}. By contrast, many countries in the Global South remain limited, resulting in a geographically uneven knowledge base that may not fully capture local priorities in health, education, and economic development \citep{Confraria_globalsouth_2017}. In terms of gender, men continue to dominate research on the SDGs overall, while women are more strongly represented in social and health-oriented topics (e.g., SDG 1 and SDG 5). These gendered patterns of participation likely reflect differences in research specialization and topic selection \citep{Nielsen_gender_2017,sugimoto2023equity}. Such disciplinary orientations may also interact with structural factors—such as differential access to funding, mentorship, and institutional leadership. Taken together, the evolution of SDG-linked psychology over the past century reveals shifting societal priorities alongside differentiated participation across countries and genders, shaping what is studied, by whom, and with what reach.

The year 2015 matters for the SDGs, but it did not mark the start of SDG study in Psychology. Many SDG-relevant problems, such as health, education, and inequality, have been studied for decades. What changed in 2015 was a turning point in attention, as it started to bundle dispersed lines of work under a common frame, making them easier to find \citep{Nilsson_policy_2016,Griggs_SDG_2013}. Our research shows a discrete shift in scholarly attention: conditional on subfield, year, and authorship characteristics, SDG-tagged psychology papers published after 2015 receive higher citations than comparable non-SDG papers, reversing a modest pre-2015 deficit. This pattern indicates that the 2015 adoption served as an exogenous coordination signal, elevating the importance and priority of SDG-aligned problems rather than simply reflecting a change in the underlying research productivity. Policy signals can sustain and broaden this attention by aligning funding and prioritizing underrepresented issues. Such a framework can be especially beneficial for under-resourced regions by aligning funding calls and fostering collaboration to build local research capacity, thereby enabling increased attention to translate into context-appropriate evidence and solutions.

Methodologically, our study uses a transparent, query-based classifier, text2sdg, to detect SDG signals from the textual corpora of titles and abstracts in scientific publications \citep{text2sdg, Wulff_SDG_2024}. Unlike approaches that require explicit SDG tags, this method infers SDG relevance even when publications do not directly reference the goals, enabling consistent coverage of the pre-2015 record. It is interpretable via term-level matches, providing auditable rationales for the lexical cues driving assignments \citep{Kahn_NLP_2022}. This, in turn, allows us to compare emphases within the same SDG and relate changes in focus to contemporaneous societal developments. The approach is readily portable to other scientific domains, bibliographic sources, and languages, as a foundation for cumulative, reproducible monitoring of SDG-relevant science \citep{Armitage_SDG_2020}.

This study also has limitations. First, our analysis is based on publications indexed in the APA databases which, although extensive, do not comprehensively represent global psychological research. Some psychology publications, particularly those in interdisciplinary journals or outside the English-speaking world, are not included in this database. Future work could incorporate additional sources, such as Scopus or Web of Science, to provide a more complete picture of psychology’s global engagement with the SDGs. Second, although text2sdg has been shown to perform well relative to other SDG detection systems (e.g., Aurora and Elsevier) \citep{Wulff_SDG_2024}, query-based methods are not perfectly accurate and can be sensitive to lexical drift. Subsequent studies could complement text2sdg with modern LLM-based approaches to obtain richer details and improve accuracy \citep[for a tutorial on LLMs, see ]{hussain2024tutorial}. Finally, while we use term-level evidence to interpret SDG assignments, we do not investigate in depth the mechanisms behind SDG studies, for example, how constructs are operationalized within the goals. Deeper validation that links text signals to study designs, measures, and outcomes would strengthen causal interpretation about the drivers and consequences of SDG-related science.

\section*{Conclusion}

Psychology has studied many of the challenges now formalized as the SDGs for more than a century, long before the UN codified them in 2015. Our analysis of 233,061 publications shows a field that has consistently mirrored societal priorities—shifting from work and education to health, inequality, and gender—but with striking blind spots. Environmental goals remain nearly absent, research production is concentrated in a small number of WEIRD countries (i.e., Western, Educated, Industrialized, Rich, and Democratic), and women are underrepresented in STEM-oriented SDGs. However, the 2015 SDG framework functioned as an effective coordination signal: SDG-labeled publications became more visible and more cited, indicating that policy cues can redirect scientific attention, and vice versa. As the 2030 deadline approaches, psychology must confront the gaps our analysis reveals—expanding engagement with climate and ecosystems, broadening global participation, and diversifying who shapes the research agenda. Whether and to what extent this is feasible---given shifting research and funding priorities among leading contributors such as the U.S. and a global autocratic turn \citep{vdem2024report}---remains far from certain. Human behavior, and the behavioral sciences that study it, are indispensable to sustainable development; whether psychology will fully realize that promise and whether societies will fully harness its evidence remain open, urgent questions. 

\section*{Data and methods}
\subsection*{APA data}
We analyzed the metadata of the full publication record of journals published by the American Psychological Association (APA), the oldest and largest publisher in Psychology. The data were provided to us by the APA. It consists of author, title, abstract, and journal information of 233,061 articles published between 1894 and 2022.  

\subsubsection*{SDG detection}
To identify SDG-related research within psychology, we applied the text2sdg package in R \citep[\href{https://www.text2sdg.io/}{www.text2sdg.io/}]{text2sdg, Wulff_SDG_2024}, which uses an ensemble of query-based labeling systems to map textual content to the 17 United Nations Sustainable Development Goals (SDGs). We constructed a text corpus by concatenating article titles and abstracts from the APA data, excluding records with missing metadata (1,074, 0.46\% of the dataset). The package detects the presence of SDG-specific keywords and assigns corresponding SDG labels to each publication. The output consists of the assigned SDGs, along with the matched query keywords that support each assignment.

\subsubsection*{Gender detection}
We inferred binary gender from first names using Genderize.io, which maps first names to gender labels with associated probabilities. Because the APA dataset lacks persistent author identifiers, we parsed the given-name field to construct ordered lists of first names for each publication. We then matched names to Genderize.io (unmatched names were left missing) and, for every publication, derived four measures: (1) female first author (Boolean; first position labeled “female”), (2) female last author (Boolean; last position labeled “female”), (3) female author share (mean of positions labeled “female”), and (4) number of authors. Coverage was high: around 90\% of articles yielded at least one name–gender match; 35\% of articles had a female first author and 32\% had a female last author. On average, articles listed 2.5 authors, with a mean female author share of 34\%.

\subsubsection*{Subfield categorization}

We identified the subfield of each APA publication by mapping each publication’s journal to the APA subject taxonomy. For journals without a direct match, we prompted GPT-4o with the APA category list and the journal name, instructing the model to return one best-fitting category. The closest APA subject based on the journal was assigned to the related publications as their subfield. 

In total, we classified 162 distinct journals across ten APA subjects. The largest shares were Clinical Psychology (31\%), Basic and Experimental Psychology (22\%), and Core of Psychology (16\%), together accounting for around 70\% of journals. The remaining categories each comprised <10\% individually: Health Psychology and Medicine (8\%), Industrial/Organizational/Management (6\%), Social Psychology and Social Processes (4\%), Educational/School/Training (4\%), Developmental Psychology (4\%), Neuroscience and Cognition (3\%), and Forensic Psychology (2\%).

\subsubsection*{Citation extraction and normalization}
We obtained citation links for articles with DOIs by querying the OpenCitations COCI API and aggregated incoming citations to a cumulative count per publication (records without DOIs were excluded). To enable comparisons across fields and years, we computed a subfield- and year-normalized citation score. Using the APA subject category (subfield) assigned to each journal and the article’s publication year, we scaled each paper’s citations by the cohort mean:
\begin{equation}\label{eq:1}
\mathrm{Norm\_Cite}_i = \frac{c_i}{\overline{c}_{f,y}},
\end{equation}
where $c_i$ is the citation count for paper $i$, and $\overline{c}_{f,y}$ is the average citation count of all papers in subfield $f$ published in year $y$. Scores greater than $1$ indicate above-average impact within the paper’s subfield and publication year.

\subsection*{Difference-in-difference in Citation analysis}
To evaluate whether SDG-related publications received greater scholarly attention, we applied a difference-in-differences (DiD) framework. Specifically, we modeled the normalized citation impact using the following regression: 

\begin{equation}\label{eq:2}
\begin{split}
\log(\text{Norm\_Cite}_i) 
= \, & \beta_0 
+ \beta_{\text{sdg}} \,\text{SDG\_paper}_i 
+ \beta_{\text{year}} \,\text{Year\_2015}_i  \\
& + \beta_{\text{DiD}\times \text{year}} 
   (\text{SDG\_paper}_i \times \text{Year\_2015}_i) \\
& + \sum_{k} \beta_k X_{ki} 
+ \epsilon_i
\end{split}
\end{equation}

where the dependent variable $\text{Norm\_Cite}_i$ is the normalized citation count of publication $i$, log-transformed to reduce skewness. The coefficients on $\text{SDG\_paper}_i$ and $\text{Year\_2015}_i$ capture baseline differences between SDG and non-SDG publications and between pre- and post-2015 periods. The coefficient $\beta_{\text{DiD}}$ captures the DiD effect, i.e., the additional change in citations for SDG-related papers after 2015 relative to non-SDG papers. The term $\sum_{k} \beta_k X_{ki}$ represents a vector of additional covariates, including gender composition of the author team (whether the first author is a woman, whether the last author is a woman, and the overall proportion of women among coauthors), the number of authors, the journal impact factor (extracted from the American Psychological Association's website \citep{APA_JIF}), and the subfield category, while $\epsilon_i$ denotes the error term.  

Based on the estimates from Equation~\ref{eq:2}, we generated predictions for SDG versus non-SDG publications before and after 2015, holding other variables constant. These predictions were back-transformed to provide expected normalized citation counts, enabling a direct comparison of whether SDG framing conferred a citation advantage across periods.


\clearpage
\bibliography{bib}


\section*{Acknowledgments}  

\paragraph*{Funding:}
Dirk U. Wulff acknowledges funding from the German Science Foundation (SPP 2317).

\paragraph*{Author contributions:}
Conceptualization, methodology, visualization, and software: X.Z., D.U.W. Data Curation: D.U.W., R.H. Formal analysis and writing—original draft: X.Z. Writing - Review \& Editing X.Z., R.H., D.U.W.
\paragraph*{Competing interests:}
There are no competing interests to declare.

\paragraph*{Data and materials availability:}
Code is available at \href{https://github.com/zxy919781142/Sustainable-Development-Goals-in-Psychology/tree/main}{github.com}. Data cannot be shared. Access to the data was granted under a signed data-use agreement, which prohibits public dissemination.
\noindent \textbf{Author contributions.} Conceptualization, methodology, visualization, and software: X.Z., D.U.W. Data Curation: D.U.W., R.H. Formal analysis and writing—original draft: X.Z. Writing - Review \& Editing X.Z., R.H., D.U.W.

\subsection*{Supplementary materials}
More details in the result\\
Model estimates\\
Figs. S1 to S2\\

\end{document}